%% file: manuscript.tex
\documentclass{IEEEcsmag}
\input{preamble.tex}

\title{Compositional Thinking\\ in Cyberphysical Systems Theory}
\author{Georgios Bakirtzis}
\affil{University of Virginia}
\author{Eswaran Subrahmanian}
\affil{Carnegie Mellon University}
\author{Cody H. Fleming}
\affil{Iowa State University}

\begin{document}
\begin{abstract}
    Engineering safe and secure cyber-physical systems
    requires system engineers to develop
    and maintain a number of model views, both dynamic and static,
    which can be seen as algebras.
    We posit that verifying the composition of requirement, behavioral, and architectural models
    using category theory gives rise 
    to a strictly \emph{compositional} interpretation of cyber-physical systems theory, which can assist in the modeling and analysis of safety-critical cyber-physical systems.
\end{abstract}
\maketitle

\chapterinitial{Applied Compositional Thinking}

   \begin{figure*}[!t]
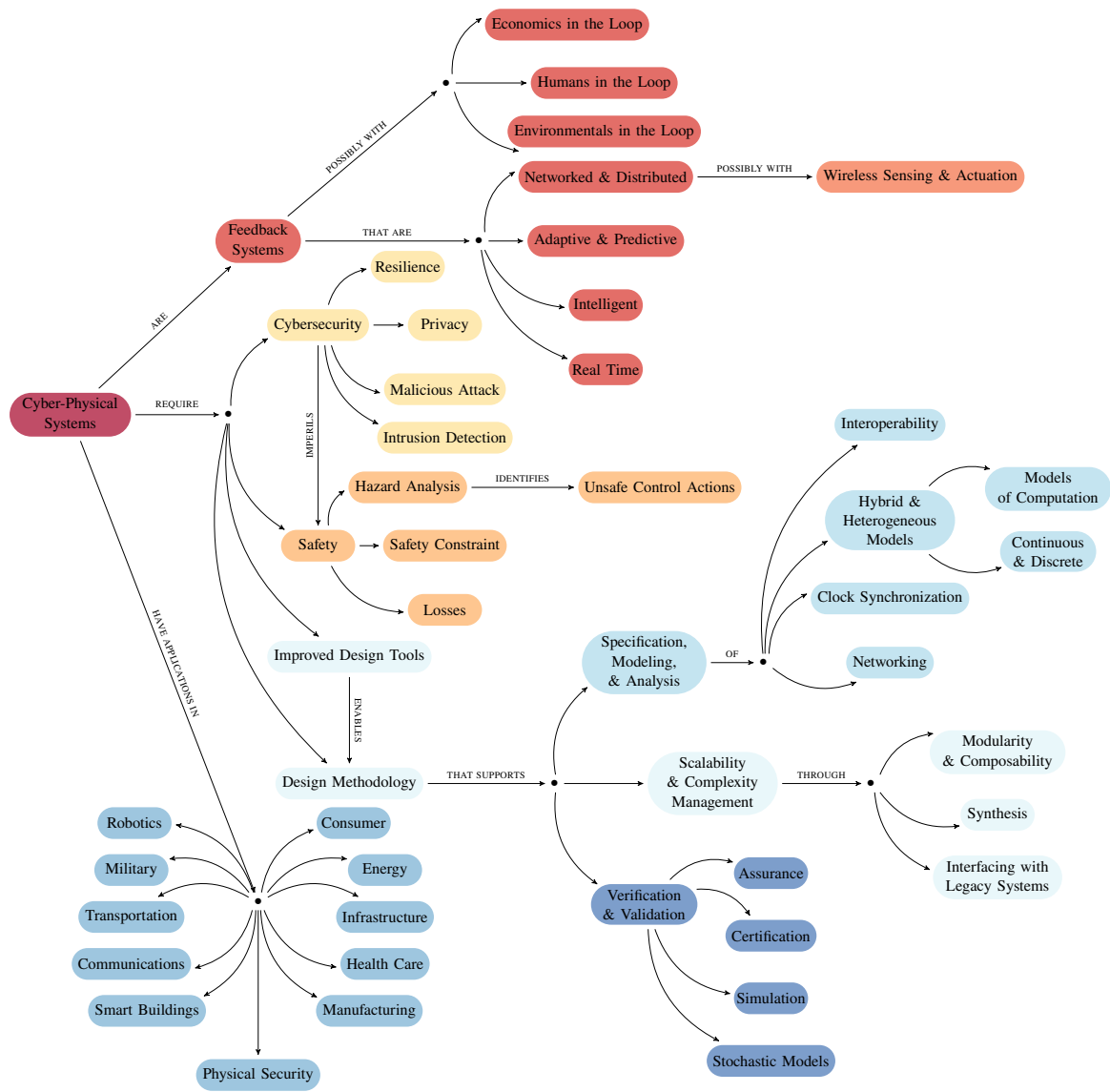

   \centering
   \includestandalone[width=1\textwidth]{figures/cps}
   \caption{A cyber-physical systems concept map showing what these systems are, what they require, and where they are being used (adapted from Asare et al.~\cite{cps}). Individual leaf nodes in this context map have been studied in detail but formal relationships between leaf node concepts are still sparse.}
   \label{fig:cps}
   \end{figure*}

Lee~\cite{lee:2006}, among others, recognized early
in the development of the field of cyber-physical systems
that there is a need for developing competing methods
to hybrid systems and process algebras.
While this is true, an important observation is
that both these formalisms form algebras.
In fact, the design
of cyber-physical systems involves the study of different algebras (Figure~\ref{fig:cps}).

There is significant research
in developing these individual algebras
and implementing composition within a particular algebra.
However, there is still an open problem about how 
to relate those paradigms that in practice represent individual models and to examine the behavior of the system as a whole must be composed too.
Compositional cyber-physical systems theory~\cite{bakirtzis:2020,bakirtzis:2020a} uses category theory
to transform data from one algebra to another and to ultimately relate them formally, such that we can compose across domains.
This provides one solution
to the open problem
of composition between formal methods
and their corresponding model views in cyber-physical system design~\cite{griffor:2017,michael:2020}.

The study of compositionality is not new
and certainly not only possible with categories.
For example, compositionality is formalized
and addressed with tools 
from control theory~\cite{alur:2018}, contract-based design~\cite{Nuzzo15b},
or monotone co-design~\cite{censi:2017}.
This is to say that we recognize that the cyber-physical systems field has had a long tradition in formalization for the study of correctness and the management of design complexity.
But the idea of category theory
in system science and formal methods is
that those approaches may be seen as operating
implicitly within a category
and making that explicit can be of use
in developing scalable and general purpose modeling tools
that can operate across categories.

Recently all these three areas of control~\cite{culbertson:2019},
contracts~\cite{bakirtzis:2020a}, and co-design~\cite[chapter~4]{fong:2019}
have been described, generalized, and unified
with fundamental notions of category theory
and mainly using the notion of functoriality -- structure preserving maps between categories
that can translate a general syntax
to the particular semantics of the application domain.
This means that we need not reconstruct the progress
that has occurred in each individual field
or model view because we can use the algebras already
in existence, but with the added benefit that we can compose between categories and, therefore, between model views.

Applied compositional thinking brings forward the practical dissemination 
of categorical results
in the field of engineering in general.
We will particularly concern ourselves 
with how research directions
in applied category theory
can be leveraged in system science and formal methods.
Composition can mean different things
both depending on the engineering field
as well as the particular context
in which composition is studied.
In cyber-physical systems the term composition can be understood as either horizontal or vertical.

Horizontal composition is relatively well studied
and refers to things like the block diagram algebra widely used
in the control community.
Vertical composition, in our definition, instead spans different domains;
a clear example in the application of systems modeling
and formal methods is how components compose
and are abstracted or refined
between requirements, behaviors, and architectures.
In order to do this right we need 
to formally relate
and verify the composition of different types
of algebras present in the design
and analysis of cyber-physical systems (the leaf nodes of Figure~\ref{fig:cps}).
At the moment, engineering researchers
and practitioners attempt
to address this issue
by developing naming spaces
and conventions. The need to capture the relationships between heterogeneous system  representations in contract-based design has led to the notion of \emph{vertical contracts}~\cite{Nuzzo18a}.
The congruence between our framework
and contract-based design is that contracts
in both senses are extensions of behavior types.
Across paradigms, as long as types agree, composition is possible.

Each of the individual formalisms
or algebras used
in the design of cyber-physical systems have had whole fields dedicated to them
with steady research progress before
and after the term cyber-physical systems came to existence.
However, the relationship between those algebras
at the moment is
either \emph{ad-hoc} 
or \emph{quasi}-formal.
Relating algebras formally can lead to new insights, methods,
and tools for the design of cyber-physical systems
that operate as we expect.

Specifically, category theory is one mathematical tool
that can verify the composition of differing views and therefore lead
to another dynamical computation systems theory
founded on the recognition
that addressing composition is important
in understand the (mis-)behavior of the system as a whole.
Category theory is the study of mathematical structures
from the perspective that, to better determine an object's purpose and behavior,
we ought to study its relationship with other objects
instead of examining the object only in itself.
Categories are intuitively congruent
with engineering cyber-physical systems
because of the existence of both dynamics
and computation in those systems,
which are modeled through a multitude of perspectives
that need to be related.

Formal methods have taken an increasingly important role
in the design of systems both
in academic~\cite{sirjani:2020} and industry~\cite{newcombe:2014} settings.
In system modeling, the SysML V2 standard is a glimpse
of the possible future of formal
and compositional methods in a practice
where the gap between requirement, behavioral,
and structural formalisms is bridged.
However, it is not clear how this relationship
will be formally defined.
This shows an increasing need
for formal and compositional methods
that verify the composition and scale system models.

While there are several applications
of category theory in engineering,
in this paper we will examine precisely this gap
between requirements, behaviors, and architectures
from the eye of compositional cyber-physical systems theory.
This is possible because composition here fundamentally gives us modularity and interoperability
within and across different types of models for free, provided we model things within the stricter paradigm of categories and algebras.
Specifically, we will show how horizontal and vertical composition
can be expressed as a formal method
using categories.

\section{Compositional Cyber-Physical Systems Theory}

\begin{figure*}[!t]
 \centering
 \subfloat[A functor maps objects.]{%
   \includegraphics[width=0.49\linewidth]{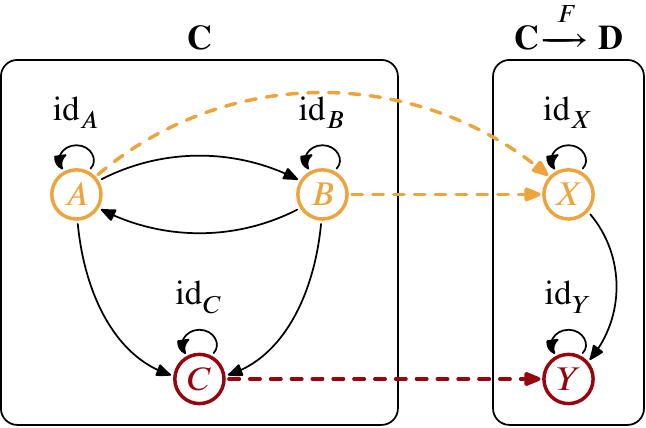}\label{fig:objects}}\hfill
 \subfloat[A functor also maps morphisms.]{%
   \includegraphics[width=0.49\linewidth]{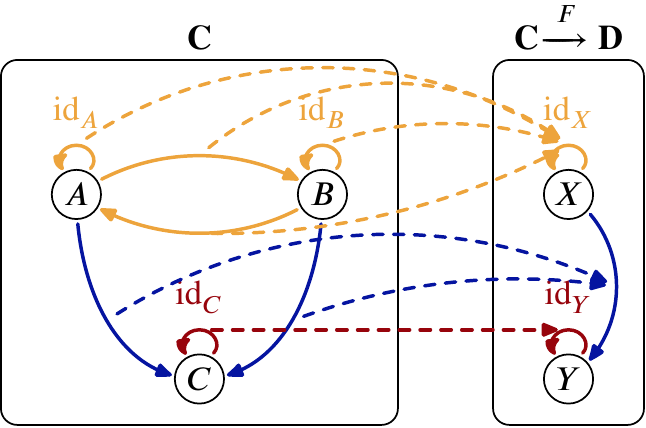}\label{fig:morphisms}}
 \caption{A functor is a structure-preserving map.}
 \label{fig:functors}
\end{figure*}

Engineers and system designers use expertise, intuition 
and inertia to decide how different analyses factor 
into higher-level, harder questions about safety, resiliency, profitability and risk. 
We can formalize (some of) that expertise 
by reconstructing existing workflows 
and best practices in categorical terms.
Though many individual features of cyber-physical systems are formal, the relationship 
between these pieces has not yet been mapped.
In general, putting systems together requires two things
per Kalman~\cite{willems:2007}:
\begin{enumerate}
    \item Getting the physics right.
    \item The rest is mathematics.
\end{enumerate}
But that sparked Willems~\cite{willems:2007}
to ask if we are using the right mathematics
to describe the science of \emph{interconnection}
and he concludes that we need to represent systems as algebras.
Category theory goes even further by composing those algebras together.

Instead of describing category theory in full, we want the reader to focus
on the categorical toolkit and refer to 
the many formal treatments on the topic, including books
such as Fong and Spivak~\cite{fong:2019}. 
However, we will give an intuitive explanation
of the fundamental concepts of category theory.

The most general notion of a category is 
that of an algebraic structure on a graph.
A category contains a collection of objects
and a collection 
of arrows (including an identity arrow)
which can formally be joined 
by a composition rule.
Examples of useful categories include the category 
of sets and functions,
of graphs and graph transformations, 
and of states and transitions between them.

The notion of functoriality between syntax and semantics we talked about previously is implemented through a structure preserving map between categories.
Functors are a predefined way of tranforming some data
and their associated interconnections to another form (figure~\ref{fig:functors}). For example, a functor can be taking the powerset of a set, where we have predefined operations on how to do that between the same category, namely the category of sets.
We can think of a functor in the application
of cyber-physical system design as preserving \emph{semantics} across the currently distinct algebras that define different interpretations
of requirements, system behaviors, and system architectures, meaning that they assign a particular algebra to a specific model view.

Often we would like to model operations happening in parallel when speaking about cyber-physical systems.
The constrained definition of a category is capable
of modeling sequential processes but not parallel ones.
A particularly useful type of category in system science is
one that is \emph{monoidal}.
Monoidal categories extend this definition 
precisely to equip with a tensor product functor, which models processes 
and/or operations that need
to happen in parallel.

As a representation of some data in a category,
see below a commutative diagram involving objects $X$, $Y$, and $Z$
identity morphism on objects, $\text{id}$, 
as well as two morphisms $f$ and $g$ 
along with their composition rule $g \circ f$.

\begin{figure*}[!t]
\centering

\includestandalone[width=\textwidth]{./figures/behavior}

\caption{A compositional model of an unmanned aerial vehicle (UAV) requires us to add extra data. The functions $\phi_\text{in}$ and $\phi_\text{out}$ create the interfaces of expected types of inputs and outputs of each box and then combine them. They are in a sense the architecture of the wiring diagram. In this case $s'$ represents the calculated state, $s$ the current state, $e$ the environment, $d$ the desired state, and $c$ the control action. This operation does not tell us what processes actually inhabit the boxes, an additional step is needed by functorially assigning behavior algebras, in this case the algebra of linear time-invariant systems $\mathcal{B}$ to each box, parallelizing the operation, and then completely determining the operation of the overall box $\text{UAV}$, as in $\mathcal{B}({L})\times \mathcal{B}({C})\times \mathcal{B}({D})\xrightarrow{\phi_{{L},{C},{D}}}
\mathcal{B}({L}\otimes{C}\otimes{D})\xrightarrow{\mathcal{B}(f)}\mathcal{B}(\text{UAV})$. Both the left and right pictures form wiring diagrams.
The left representation is better suited
for incorporating this diagrammatic reasoning within modeling tools.}
\label{fig:uav}
\end{figure*}

\[
\begin{tikzcd}[ampersand replacement=\&]
X\arrow[r,"f"]\arrow[dr,"g\circ f"'] \arrow[loop left,"\text{id}_X"] \& Y\arrow[d,"g"] \arrow[out=65,in=25,loop,"\text{id}_Y"] \\
\& Z \arrow[loop right,"\text{id}_Z"]
\end{tikzcd}
\]

The objects and morphisms can take several forms, the most familiar
of which is perhaps the category of sets and functions,
where objects are sets and morphisms are functions and composition is function composition.
The example of a category we will use
to show the utility of category theory
in engineering is the labeled boxes and wiring diagrams category~\cite{schultz:2020},
where the objects are empty placeholders for processes,
and the morphisms are functions
or relations that construct input
and output interfaces for each labeled box.

By having the mathematical structure
of a category it is now possible
to translate between model views
as long as they form categories 
and their behavior can be represented
as an algebra.
This is achieved by defining 
 a precise way, that can also be implemented algorithmically, of translating the objects from one category to another (Figure~\ref{fig:objects})
and similarly for the morphisms (Figure~\ref{fig:morphisms}).
An example 
of a functorial operation
between categories of sets is transforming a set
to its power set.
A \emph{monoidal} functor comes with comparison morphisms that involve the tensor product,
allowing us to use the same idea
in the setting of parallel operations.
This might seem like new and confusing jargon
for those unfamiliar with category theory
but once it is understood it allows us
to summarize a vast number of mathematical techniques
as well as formal methods and system models
by allowing us to assign algebras that contain parallel processes.

We use these algebraic structures
to describe the syntax and semantics
for cyber-physical systems compositionally.
We show how a wiring diagram has an architecture;
that is, how interfaces are constructed,
and a behavior; that is, how the notion
of functorial semantics takes the architecture -- the syntax --
and assigns multiple semantics, in this case of linear time-invariant system models.

In other cases we could use the same structures
to assign other models of cyber-physical system behavior,
such as labelled transitioned systems or Moore machines.
Hierarchical decompositions are also necessary to construct the desired vertical composition structure.
To achieves this, we are going to use the \emph{slice} category
of a wiring diagram to decompose the model
to a particular system architecture -- the hardware composition
of the embedded system that implements the behaviors we have assigned,
particularly for a model of an unmanned aerial vehicle (UAV).
Finally, we will restrict the behavior
of the system using a categorical description
of contract-based design for behavior. In this respect, we regard contracts as representations of the system requirements.
However, note that those requirements operate
both as restrictions in behavioral and architectural system models.

The general principle of applied compositional thinking is the following.
\begin{enumerate}
    \item Wiring diagrams define how things ought to be connected and in what way.
    \item Algebras assign a particular formalism or behavior to the empty processes (boxes) inhabiting the wiring diagram.
\end{enumerate}

Wiring diagrams are the theory of the class of systems we want to model but only insofar as they are assigned a behavioral \emph{model} through an algebra (monoidal functor).
The implication of this statement is that for each wiring diagram, we can define
 many algebras and, therefore, can capture -- in the same arrangement of components -- a number of formalisms
as well as translate between formalisms.
Additionally, any algebra is a monoidal functor from the category of wiring diagrams to the category of sets.
To deal with the disparate leaf nodes (Figure~\ref{fig:cps})
that are needed for the design of a complete cyber-physical system it is useful to work in higher levels of abstraction, for example, topological spaces, graphs, vector bundles,
but it is crucial to translate to a realizable design,
which usually means that even if we do not do the original operations in the category of sets, the result should be translatable to the category of sets.

\subsection{Horizontal Composition}

As an example we model an unmanned aerial vehicle compositionally (Figure~\ref{fig:uav}).
There are two observations stemming
from our UAV example:
\begin{enumerate}
    \item The functions $\phi_\text{in}$ and $\phi_\text{out}$ create the interfaces of input-output relationships that completely determine the block UAV.
    \item But by knowing the types and values expected at interfaces we do not really know \emph{how} the system transforms this type of data.
    We, therefore, need an algebra $\mathcal{B}$ that assigns precisely the behavior of linear time-invariant systems.
\end{enumerate}

A result of doing this categorically
is the realization that extra information is often necessary,
which is often omitted or glossed over by other composition operators and methods.
In order for the system to compose nicely it is paramount that the \emph{type} of system residing within all boxes in a wiring diagram are either the same, for example, Moore machines inhabit all boxes, or that we know the results of the composition of two types, for example, ordinary differential equations and Moore machines.

\begin{figure*}[!t]
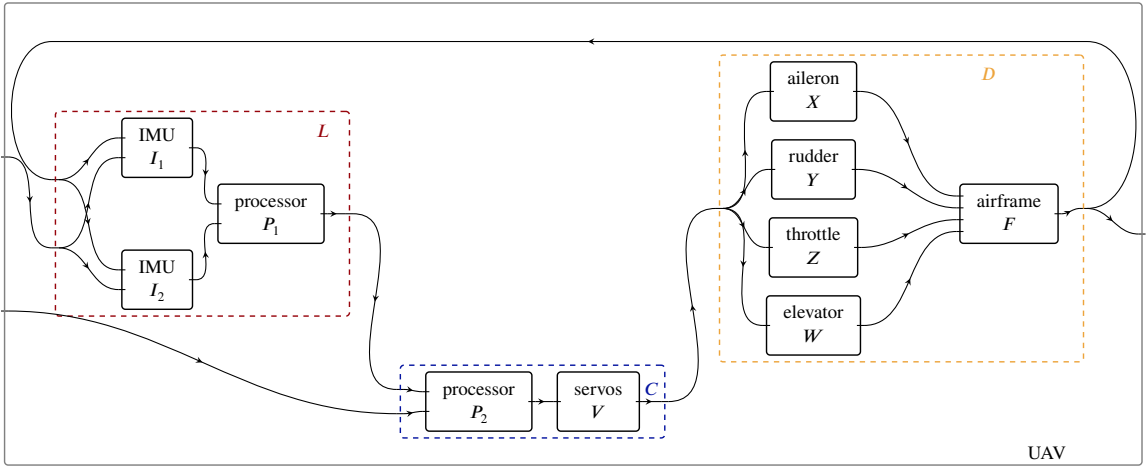

\centering

\includestandalone[width=\textwidth]{./figures/architecture}
\caption{The architecture of the system resides within the slice category of our initial system behavior diagram (Figure~\ref{fig:uav}) producing a formal trace between the two, thereby allowing us to perform more analyses formally.
While this decomposition focuses on the one possible implementation of the embedded system aspects, this is only because it was modeled by a computer engineer. An aerospace engineer could augment this model with further decompositions of the airframe or motors as long as they got the physics right.
}
\label{fig:architecture}
\end{figure*}

This means that in our example, all boxes must take the form of a linear time-invariant system of the standard form, even though in the block diagram algebra used in control describes the totality of the control system behavior.
Adding this extra information is important but the solution is straightforward: reduce sensor and controller to trivial functions because they will compose to the form of the linear time-invariant system representing the dynamics we expect (Figure~\ref{fig:uav}).
The main benefit of implementing
this algebra machinery is
that it uniquely determines the description
of the composite system
in terms of its subsystems
and their interconnections.

One way of assigning requirements to this behavior is
to translate them 
into contracts~\cite{benveniste:2018,filippidis:2018}.
As with behavior using the algebra of contract we can first think
of contracts as a relation of inputs and outputs
of each box in the wiring diagram.

Then to each complex arrangement
of boxes assign the following pullback,
which given subsystems contracts
produces a composite contract
of the system model.

\[
\begin{tikzcd}[ampersand replacement=\&,column sep=.25in, row sep=.35in]
\&\bullet\ar[rr]\ar[d,>->]\ar[dl,two heads,bend right] \&\& {\color{effectivenessCurve}R_X} \ar[d,effectivenessCurve,>->] \\
{\color{CostCurve}R_Y}\ar[dr,>->,CostCurve,bend right] \&\inp{Y}\times\out{X}\ar[rr,"{(\inp{\phi},\pi_2)}"']\ar[d,"\text{id} \times \out{\phi}"description] \&\& 
{\color{effectivenessCurve}\inp{X}\times\out{X}} \\
\& {\color{CostCurve}\inp{Y}\times\out{Y}} \&\&
\end{tikzcd}
\]

For a restriction in each subsystem in the composite we can represent that restriction as
a set of allowable pairs that can be observed as inputs and outputs and from this formulation we will always be able to
calculate the allowable input/output pairs for the composite system in terms of the specific wiring given by the architecture
of the wiring diagram.

Horizontal composition is defined
as the interconnection of operations in one model type.
The two step process is to use the theory
of interconnection to define the architecture
of the base wiring diagram
and then inhabit the boxes
of this arrangement with a particular behavioral model~\cite{bakirtzis:2020a}.
This brings our compositional model
into the same realm as behavioral diagrams in SysML
but with the possibility
of simulated behavior as in MATLAB Simulink.
The main benefit of this approach is that we can assign a number of algebras
by describing models in categories.
In this case, we assigned part
of our requirements using the contracts algebra
over the same model.

\subsection{Vertical Decomposition}

There are cases where the behavioral interpretation
of the system is not sufficient
to examine certain properties about a system.
One such case is security, where one approach
to finding vulnerabilities is to traverse a graph
of the expected or implemented system architecture~\cite{bakirtzis:2020b}.
However, those approaches are static
and often are behaviorally unaware,
meaning that there is no way to know what behavior is being affected
by a particular successful exploit.
This is not necessarily an issue in information technology (IT) systems
but in the case of cyber-physical systems, where the dynamics are richer it is necessary to know what physical behaviors the exploit might affect.

One way of understanding what a slice category is
 in a given wiring diagram is to think about it as splitting. We split or open up morphisms (the arrows) within the wiring diagram category
in order to allow a higher degree of refinement.
We can think of this as working
with hyper graphs but with the extra structure
of a category.
Visually this morphism split is opening up the model of system behavioral (Figure~\ref{fig:uav})
to a model of system architecture (Figure~\ref{fig:architecture}).
The splitting is a form of refinement, but by using the slice category it forms a formal trace between behaviors and architectures.

Vertical decomposition not only refers to decomposing system models, but at the same time it can mean the relation of different algebras, in this case, for
example, the algebra of behavior with the algebra of contracts.
A similar argument could be made
for applying contracts
to the eventual architecture
and its associated algebra (with its own corresponding architecture
and behavior in the wiring diagram representation).

To understand the disjointedness
of the current modeling process we can think about how we might mathematically assign behaviors
and restrictions to an empty process.
To the same wiring diagram (a collection of empty processes
and wired between them) we can assign different formalisms.
This is the reason for studying empty processes, namely that it gives us a way of decoupling syntax from semantics.

\[
\begin{tikzcd}[row sep=.05in,column sep=.7in,ampersand replacement=\&]
\& \mathcal{B}(X) \ar[dd,"\mathcal{B}(f)"] \\
\& \phantom{A} \\
X\ar[dd,"f"'] \& \mathcal{B}(Y) \\
\phantom{B}\ar[uur,bend left,dotted] \ar[ddr,bend right,dotted] \& \\
Y \& \mathcal{C}(X) \ar[dd,"\mathcal{C}(f)"] \\
\& \phantom{C} \\
\& \mathcal{C}(Y)
\end{tikzcd}
\]

This represents the way a system designers might approach developing contracts
over an already known behavior.
We would instead prefer a more formal relationship of these two algebras.
In the categorical case we have been explicit
in assigning behavior and contracts
in the form of algebras.
This allows us to use a natural transformation
-- a map between functors -- to reflect changes from the behavioral paradigm to the contracts paradigm,
which can be seen as the relationship
between (distinct) algebraic representations, precisely as the behavior algebra ($\mathcal{B}$) and contracts algebra ($\mathcal{C})$ in our work.

\[
    	\scalebox{0.75}{
           \tdplotsetmaincoords{70}{125}
           \begin{tikzpicture}[tdplot_main_coords, scale=0.4]

           \begin{scope}[canvas is xz plane at y=0]
           \draw[thick] (0,0) ellipse (6cm and 8cm);

           \draw[fill] (0.5,5) circle (3pt) node[align=left, black, left] {$FA$};
           \node (f1) at (0.5,5) {};
           \draw[fill] (-2,0) circle (3pt) node[align=center, black, below right] {$FB$};
           \node (g1) at (-2,0) {};
           \draw[fill] (1.5,-4) circle (3pt) node[align=right, black, left] {$FC$};
           \node (h1) at (1.5,-4) {};

       \draw[->, >=stealth, shorten >=.03cm] (f1) to node[right, black] {$Ff$} (g1);
           \draw[->, >=stealth, shorten >=.03cm] (g1) to node[right, black] {$Fg$} (h1);
       \draw[->, >=stealth, shorten >=.03cm] (h1) to node[left, black] {$Fh$} (f1);

           \end{scope}

           \begin{scope}[canvas is xz plane at y=12]
           \draw[thick] (0,0) ellipse (6cm and 8cm);

           \draw[fill] (0.5,5) circle (3pt) node[align=left, black, right] {$GA$};
           \node (f2) at (0.5,5) {};
           \draw[fill] (-2,0) circle (3pt) node[align=center, black, below right] {$GB$};
           \node (g2) at (-2,0) {};
           \draw[fill] (1.5,-4) circle (3pt) node[align=right, black, below] {$GC$};
           \node (h2) at (1.5,-4) {};

           \draw[->, >=stealth, shorten >=.03cm] (f2) to node[right, black] {$Gf$} (g2);
           \draw[->, >=stealth, shorten >=.03cm] (g2) to node[right, black] {$Gg$} (h2);
       \draw[->, >=stealth, shorten >=.03cm] (h2) to node[left, black] {$Gh$} (f2);

           \draw[->, >=stealth, shorten >=.03cm, thick] (f1) to node[above, black]{$\eta_A$} (f2);
           \draw[->, >=stealth, shorten >=.03cm, thick] (g1) to node[above, black]{$\eta_B$} (g2);
           \draw[->, >=stealth, shorten >=.03cm, thick] (h1) to node[above, black]{$\eta_C$} (h2);
           \end{scope}
           \end{tikzpicture}}
\]

Assume that we assign
to an empty process a particular formalism of behavior and contracts in the form of algebras.
But, then, as it so often happens within design a change must occur in the behavior, a range of outputs from one box has changed.
By having a clearly defined natural transformation -- that moreover behaves well with the parallel operation -- it is possible to reflect what that means in the contracts we have defined for the same system model.
By using the above relationship between functors; that is, the natural transformation,
we are doing precisely that.

\[
\begin{tikzcd}[ampersand replacement=\&]
\mathcal{B}(X) \ar[d,"\mathcal{B}(f)"'] \ar[r,"\alpha_X"] \& \mathcal{C}(X)\ar[d,"\mathcal{C}(f)"] \\
\mathcal{B}(Y)\ar[r,"\alpha_Y"'] \& \mathcal{C}(Y)
\end{tikzcd}
\]

The same argument has also implications
for scalability.
Knowing when a change in one modeling paradigm
causes something else to be changed, augmented, or rethought
within a modeling tool could reduce errors and increase efficiency in the system modeling process.
However, while we can make a clear mathematical argument for verified composition, scalability requires that we build the tools
and test our assumptions with practicioners in the field.

The promise of category theory is not that it will change the mathematics
of a particular application field but, rather,
that it will organize the information such that there might be new insights
and techniques associated with translating between data (\emph{verified composition})
and managing complexity (\emph{scalability}).

\section{Category Theory for the Engineer}

Compositionality is one
of the open problems in cyber-physical system design,
as is argued in the NIST cyber-physical system framework~\cite[pages 11-12]{griffor:2017}.
In general, we perceive
that categorical models
in engineering are currently rather unfledged
and category theory is a ripe field
with developed mathematical tools
that can be transitioned
to formal methods in the design
of cyber-physical systems.
The main benefits of applied compositional thinking is verified composition
and scalability of different system models.

As practitioners of formal methods, we would like to know
if our contracts or linear temporal specifications agree
with our linear time-invariant system or labelled transition state space models.
Therefore, the problem in cyber-physical systems is not only contained
on which algebra to use to examine or assure a particular property
of the system but also how to relate those algebras.
The importance of this perspective is
that algebras, even implicitly, exist
at any given level of system design
and their composition gives rise
to well-defined traces in models.

Further, as we move throughout the lifecycle, system architecture models
must be developed and tested for behavioral bisimilarity
with the models we construct in the early concept design phase,
for example, to assure that our safety analysis holds.
Additionally, we might at that point want
to check qualities that heavily depend upon the composition
of the system in hardware and software, such as security \cite{bakirtzis2021yoneda}.
Additionally, even later in the lifecycle we want
to test our models with data harvested
from the designed and deployed system.
Here too, the same algebraic structures can be extended
by this data gathered in the field.

One way to partially combat the increasing 
but necessary complexity
of those systems
we need to make sure that all these algebras are traceable.
By being traceable we are therefore able
to apply several lightweight formal methods
within one overarching model.
The recent development of algebraic software based
on category theory can assist us
in developing modeling tools.
By developing compositional modeling tools
we can manage and make formal all diverse views for design,
thereby bringing to practice all this theoretical work.

\bibliographystyle{IEEEtran}
\balance
\bibliography{manuscript}

\end{document}

%% file: preamble.tex
\usepackage[utf8]{inputenc}
\usepackage[colorlinks,urlcolor=blue,linkcolor=blue,citecolor=blue]{hyperref}
\usepackage{amsmath}
\usepackage{stix}
\usepackage{standalone}
\usepackage{subcaption}
\usepackage{tikz}
\usetikzlibrary{shapes, shapes.geometric,arrows, backgrounds, calc, hobby, positioning,cd}
\usepackage{tikz-3dplot}
\usepackage{balance}

\usepackage{xcolor}
\definecolor{cps}{RGB}{165,0,38}
\definecolor{feed}{RGB}{215,48,39}
\definecolor{wireless}{RGB}{244,109,67}
\definecolor{safety}{RGB}{253,174,97}
\definecolor{sec}{RGB}{254,224,144}
\definecolor{method}{RGB}{224,243,248}
\definecolor{spec}{RGB}{171,217,233}
\definecolor{app}{RGB}{116,173,209}
\definecolor{vv}{RGB}{69,117,180}
\definecolor{yel}{RGB}{237, 164, 60}

\definecolor{CostCurve}{RGB}{5, 20, 160}
\definecolor{effectivenessCurve}{RGB}{151, 4, 12}

\newcommand{\inp}[1]{{#1}_\mathrm{in}}
\newcommand{\out}[1]{{#1}_\mathrm{out}}

\tikzset{vertex style/.style={
    fill=#1!70,
    text=black,
    rounded rectangle,
    minimum width=2cm,
    minimum height=0.75cm,
    font=\normalsize,
    outer sep=3pt, %
  },
}

\tikzset{
  text style/.style={
    sloped, %
    text=black,
    font=\footnotesize,
    above
  }
}

\tikzcdset{%
   arrow style=tikz,
   diagrams={>={Classical TikZ Rightarrow[angle=63:4pt, line width=.6pt]}},
   arrows={semithick}
}

\usetikzlibrary{positioning,decorations.pathreplacing,decorations.markings,arrows.meta,calc,fit,quotes,cd,math,arrows,backgrounds,shapes.geometric}

\tikzcdset{%
   arrow style=tikz,
   diagrams={>={Classical TikZ Rightarrow[angle=63:4pt, line width=.6pt]}},
   arrows={semithick}
}

\tikzset{
  tick/.style={postaction={
    decorate,
    decoration={markings, mark=at position 0.5 with {\draw[-] (0,.4ex) -- (0,-.4ex);}}}
  },
  tickx/.style={
    postaction={ decorate,
      decoration={markings,
        mark=at position 0.5 with {
          \fill circle [radius=.28ex];
        }
      }
    }
  }
}

\tikzset{
   dom/.style={append after command={coordinate[alias=dom#1]}},
   domA/.style={dom=A}, domB/.style={dom=B},
   domC/.style={dom=C}, domD/.style={dom=D},
   domE/.style={dom=E}, domF/.style={dom=F},
   cod/.style={append after command={coordinate[alias=cod#1]}},
   codA/.style={cod=A}, codB/.style={cod=B},
   codC/.style={cod=C}, codD/.style={cod=D},
   codE/.style={cod=E}, codF/.style={cod=F}
}

\tikzset{
   oriented WD/.style={%
      every to/.style={out=0,in=180,draw},
      label/.style={
         font=\everymath\expandafter{\the\everymath\scriptstyle},
         inner sep=0pt,
         node distance=2pt and -2pt},
      semithick,
      node distance=1 and 1,
      decoration={markings, mark=at position .5 with {\arrow{stealth};}},
      ar/.style={postaction={decorate}},
      execute at begin picture={\tikzset{
         x=\bbx, y=\bby,
         every fit/.style={inner xsep=\bbx, inner ysep=\bby}}}
      },
   bbx/.store in=\bbx,
   bbx = 1.5cm,
   bby/.store in=\bby,
   bby = 1.75ex,
   bb port sep/.store in=\bbportsep,
   bb port sep=2,
   bb port length/.store in=\bbportlen,
   bb port length=4pt,
   bb min width/.store in=\bbminwidth,
   bb min width=1cm,
   bb rounded corners/.store in=\bbcorners,
   bb rounded corners=2pt,
   bb small/.style={bb port sep=1, bb port length=2.5pt, bbx=.4cm, bb min width=.4cm, bby=.7ex},
   bb Small/.style={bb port sep=1, bb port length=2.5pt, bbx=.5cm, bb min width=.5cm, bby=1ex},
   bb/.code 2 args={%
      \pgfmathsetlengthmacro{\bbheight}{\bbportsep * (max(#1,#2)+1) * \bby}
      \pgfkeysalso{draw,minimum height=\bbheight,minimum width=\bbminwidth,outer sep=0pt,
         rounded corners=\bbcorners,thick,
         prefix after command={\pgfextra{\let\fixname\tikzlastnode}},
         append after command={\pgfextra{\draw
            \ifnum #1=0{} \else foreach \i in {1,...,#1} {
               ($(\fixname.north west)!{\i/(#1+1)}!(\fixname.south west)$) +(-\bbportlen,0) coordinate
               (\fixname_in\i) -- +(\bbportlen,0) coordinate (\fixname_in\i')}\fi %
            \ifnum #2=0{} \else foreach \i in {1,...,#2} {
               ($(\fixname.north east)!{\i/(#2+1)}!(\fixname.south east)$) +(-\bbportlen,0) coordinate
               (\fixname_out\i') -- +(\bbportlen,0) coordinate (\fixname_out\i)}\fi;
         }}}
   },
   bb name/.style={append after command={\pgfextra{\node[anchor=north] at (\fixname.north) {#1};}}}
}

%% file: figures/cps.tex
\begin{tikzpicture}[node distance=3cm and 6cm,>=stealth']
 \node[vertex style=cps, align=center] (cps) {Cyber-Physical\\ Systems};

 \node[vertex style=feed, above of=cps,xshift=12em, yshift=3em, align = center] (fb) {Feedback\\ Systems}
 edge [<-] node[text style,above]{\textsc{are}} (cps);
 
 \node[above of=fb, xshift=12em, yshift=2em] (fb-bullet) {•}
 edge [<-] node[text style, above]{\textsc{possibly with}} (fb);
 \node[vertex style=feed,above right of=fb-bullet, yshift=-2em, xshift=2em] (econ) {Economics in the Loop}
 edge [<-, bend right] (fb-bullet);
 \node[vertex style=feed, right of=fb-bullet, xshift=2em] (human) {Humans in the Loop}
 edge [<-] (fb-bullet);
 \node[vertex style=feed, below of=human, yshift=5em] (env) {Environmentals in the Loop}
 edge [<-, bend left] (fb-bullet);
 
  \node[right of=fb, xshift=6em] (fb-bullet2) {•}
  edge [<-] node[text style, above]{\textsc{that are}} (fb);
  \node[vertex style = feed, right of=fb-bullet2] (ad) {Adaptive \& Predictive}
  edge [<-] (fb-bullet2);
  \node[vertex style = feed, above of = ad, yshift=-4em] (network) {Networked \& Distributed}
  edge [<-, bend right] (fb-bullet2);
  \node[vertex style = wireless, right of = network, xshift=12em] {Wireless Sensing \& Actuation}
    edge [<-] node[text style, above]{\textsc{possibly with}} (network);
  \node[vertex style = feed, below of = ad, yshift = 4em] (intel) {Intelligent}
  edge[<-, bend left] (fb-bullet2);
    \node[vertex style = feed, below of = intel, yshift = 4em] {Real Time}
  edge[<-, bend left] (fb-bullet2);
  
\node[right of = cps, xshift=2em] (req-bullet) {•}
 edge [<-] node[text style, above]{\textsc{require}} (cps);
 \node[vertex style=sec, above right of = req-bullet] (sec) {Cybersecurity}
 edge [<-, bend right] (req-bullet);
    \node[vertex style=sec, above right of = sec, yshift=-2em] {Resilience}
     edge [<-, bend right] (sec);
    \node[vertex style=sec, right of = sec]  (priv) {Privacy}
     edge [<-] (sec);
    \node[vertex style=sec, below of = priv, yshift=4em]  (ma) {Malicious Attack}
     edge [<-, bend left] (sec);
    \node[vertex style=sec, below of = ma, yshift=5em]  {Intrusion Detection}
     edge [<-, bend left] (sec);
 \node[vertex style=safety, below of = sec, yshift=-6em] (safety) {Safety}
 edge [<-, bend left] (req-bullet)
 edge [<-] node[text style, below, rotate=180, yshift=1.1em]{\textsc{imperils}} (sec);
    \node[vertex style=safety, above right of = safety, yshift=-2em] (hazop) {Hazard Analysis}
     edge [<-, bend right] (safety);
        \node[vertex style=safety, right of = hazop, xshift=8em] {Unsafe Control Actions}
        edge [<-] node[text style, above]{\textsc{identifies}} (hazop);
    \node[vertex style=safety, right of = safety] (sc) {Safety Constraint}
     edge [<-] (safety);
    \node[vertex style=safety, below of = sc, yshift=4em] {Losses}
     edge [<-, bend left] (safety);
 
  \node[vertex style=method, below of = safety, xshift=2em, yshift=1em] (tools) {Improved Design Tools}
 edge [<-, bend left] (req-bullet);
 
 \node[vertex style=method, below of = tools] (method) {Design Methodology}
 edge [<-, bend left] (req-bullet)
 
 edge [<-] node[text style, above] {\textsc{enables}} (tools);

\node[right of = method, xshift=5em] (method-bullet) {•}
 edge [<-] node[text style, above] {\textsc{that supports}} (method);
 \node[vertex style=spec, above right of = method-bullet, align=center, yshift=2em] (spec) {Specification,\\
 Modeling,\\
 \& Analysis}
 edge [<-, bend right] (method-bullet);
    \node[right of = spec, xshift=-0.5em] (spec-bullet) {•}
    edge [<-] node[text style, above] {\textsc{of}} (spec);
        \node[vertex style = spec, right of = spec-bullet] (networking) {Networking}
        edge [<-, bend left] (spec-bullet);
        \node[vertex style = spec, above of = networking, align=center,yshift=1em] (hybrid) {Hybrid \&\\ Heterogeneous\\ Models}
        edge [<-, bend right] (spec-bullet);
            \node[vertex style = spec, right of = hybrid, align=center, yshift=-2em, xshift=2em] {Continuous\\ \& Discrete}
            edge [<-, bend left] (hybrid);
            \node[vertex style = spec, right of = hybrid, align=center, yshift=2em, xshift=2em] {Models\\ of Computation}
            edge [<-, bend right] (hybrid);
        \node[vertex style = spec, above of = hybrid, align=center, yshift=-2em] (inter) {Interoperability}
        edge [<-, bend right] (spec-bullet);
        \node[vertex style = spec, above of = networking, align=center, yshift=-4em]  {Clock Synchronization}
        edge [<-, bend right] (spec-bullet);     
 \node[vertex style=method, right of = method-bullet, align=center, xshift=2em] (comp) {Scalability\\ \& Complexity\\ Management}
 edge [<-] (method-bullet);
     \node[right of = comp, xshift=2em] (comp-bullet) {•}
     edge [<-] node[text style, above] {\textsc{through}} (comp);
        \node[vertex style = method, right of = comp-bullet, align=center, yshift=2em] {Modularity\\ \& Composability}
        edge [<-, bend right] (comp-bullet);
        \node[vertex style = method, right of = comp-bullet, align=center, yshift=-2em] {Synthesis}
        edge [<-, bend left] (comp-bullet);
         \node[vertex style = method, right of = comp-bullet, align=center, yshift=-6em] {Interfacing with\\ Legacy Systems}
        edge [<-, bend left] (comp-bullet);
 \node[vertex style=vv, below right of = method-bullet, align=center, yshift=-2em] (vv) {Verification\\ \& Validation}
 edge [<-, bend left] (method-bullet);
        \node[vertex style = vv, right of = vv, align=center, yshift=2em] {Assurance}
        edge [<-, bend right] (vv);
        \node[vertex style = vv, right of = vv, align=center, yshift=-2em] {Certification}
        edge [<-, bend right] (vv);
        \node[vertex style = vv, right of = vv, align=center, yshift=-6em] {Simulation}
        edge [<-, bend left] (vv);
        \node[vertex style = vv, right of = vv, align=center, yshift=-10em] {Stochastic Models}
        edge [<-, bend left] (vv);

 \node[below of = cps, xshift=12em, yshift=-23em] (app-bullet) {•}
 edge [<-] node[text style, above]{\textsc{have applications in}} (cps);
        \node[vertex style = app, left of = app-bullet, align=center, yshift=2em] {Military}
        edge [<-, bend left] (app-bullet);
        \node[vertex style = app, right of = app-bullet, align=center, yshift=5em, xshift=-2em] {Consumer}
        edge [<-, bend right] (app-bullet);
        \node[vertex style = app, right of = app-bullet, align=center, yshift=2em] {Energy}
        edge [<-, bend right] (app-bullet);
        \node[vertex style = app, right of = app-bullet, align=center, yshift=-1em] {Infrastructure}
        edge [<-, bend right] (app-bullet);
        \node[vertex style = app, right of = app-bullet, align=center, yshift=-4em] {Health Care}
        edge [<-, bend left] (app-bullet);
        \node[vertex style = app, right of = app-bullet, align=center, yshift=-7em, xshift=-1em] {Manufacturing}
        edge [<-, bend left] (app-bullet);
        \node[vertex style = app, left of = app-bullet, align=center, yshift=5em, xshift=.2em] {Robotics}
        edge [<-, bend left] (app-bullet);
        \node[vertex style = app, below of = app-bullet, align=center, yshift=-3em] {Physical Security}
        edge [<-] (app-bullet);
        \node[vertex style = app, left of = app-bullet, align=center, yshift=-7em, xshift=1em] {Smart Buildings}
        edge [<-, bend right] (app-bullet);
        \node[vertex style = app, left of = app-bullet, align=center, yshift=-4em] {Communications}
        edge [<-, bend right] (app-bullet);
        \node[vertex style = app, left of = app-bullet, align=center, yshift=-1em] {Transportation}
        edge [<-, bend left] (app-bullet);
\end{tikzpicture}

%% file: figures/behavior.tex
\begin{tikzpicture}[oriented WD, bbx = .5cm, bby =.8ex, bb min width=.5cm, bb port length=2pt, bb port sep=1]
\node[bb={2}{1}] (L) {\begin{tabular}{c}
     sensor  \\
     ${L}$
\end{tabular}};
\node[bb={2}{1}, below right= 0.01cm and 1cm of L] (C) {
\begin{tabular}{c}
     controller\\ ${C}$
\end{tabular}};
\node[bb={1}{1}, above right= 0.01cm and 1cm of C] (D) { \begin{tabular}{c}
     dynamics\\ ${D}$
\end{tabular}};
\node[bb={2}{1}, fit={($(L.west)-(1,0)$) ($(D.east)+(1,0)$) ($(L.north)+(0,5)$) ($(C.south)-(0,3)$)},gray] (A) {};
 \draw[ar] (A_in1') to (L_in2);
  \draw[ar] (A_in2') to (C_in2);
 \draw[ar] (L_out1) to (C_in1);
 \draw[ar] (C_out1) to (D_in1);
 \draw[ar] (D_out1) to (A_out1');
  \draw[ar] let \p1=(D.north east), \p2=(L.north west), \n1={\y1+2*\bby},  \n2=\bbportlen
  in (D_out1) to [in=0] (\x1-\n2,\n1) -- (\x2-\n2,\n1) to[out=180] (L_in1);
 \node (text) [below=8 of C,rectangle,align=left]
{$ \phi_\mathrm{in} \colon\mathbb{R}^3\times\mathbb{R}^2\to\mathbb{R}^5,\; (s',c,s,e,d)\mapsto(s,e,s',d,c)$ \\
$ \phi_\mathrm{out} \colon\mathbb{R}^3\to \mathbb{R},\;(s',c,s)\mapsto s$};
\node[bb={2}{1},right=6 of D] (L') {$L$}; 
\node[bb={2}{1},below=1 of L'] (C') {$C$};
\node[bb={1}{1},below=1 of C'] (D') {$D$};
\node[dashed,fit=(L')(D'),draw,inner sep=2] {};
\node[bb={2}{1},fit=(L')(D'),inner ysep=23, inner xsep=26] (A') {};
\draw[ar] (A'_in1') to (L'_in2);
  \draw[ar] (A'_in2') to (C'_in2);
 \draw[ar] let \p1=(L'.north east), \p2=(L'.north west), \n1={\y1+2*\bby},\n2=\bbportlen in (L'_out1) to [in=0] (\x1+\n2,\n1) -- (\x2-\n2,\n1) to [out=180] (C'_in1);
\draw[ar] (D'_out1) to (A'_out1');
\draw[ar] let \p1=(L'.north east), \p2=(L'.north west), \n1={\y1+4*\bby},  \n2=\bbportlen in (D'_out1) to [in=0] (\x1+\n2,\n1) -- (\x2-\n2,\n1) to[out=180] (L'_in1);
 \draw[ar] let \p1=(D'.south east), \p2=(D'.south west), \n1={\y1-2*\bby},\n2=\bbportlen in (C'_out1) to [in=0] (\x1+\n2,\n1) -- (\x2-\n2,\n1) to [out=180] (D'_in1);
\draw[label]
node[left=.1 of A_in1] {$\mathbb{R}$}
node[left=.1 of A_in2]{$\mathbb{R}$}
node[right=.1 of A_out1]{$\mathbb{R}$}
node[above right= 7 and 3 of L_out1]{$\mathbb{R}$}
node[above right= 3 and 1 of C_out1]{$\mathbb{R}$}
node[below right= 1 and 1 of L_out1]{$\mathbb{R}$}
node[above right= of A_in1']{$e$}
node[above right= of A_in2']{$d$}
node[above right= of L_out1]{$s'$}
node[below right= of C_out1]{$c$}
node[below right= of D_out1]{$s$}
node[below right= 1 and 6 of C]{UAV}
node[above left= of A'_in1]{$e$}
node[above left= of A'_in2]{$d$}
node[right=.2 of L'_out1]{$s'$}
node[right=.2 of C'_out1]{$c$}
node[above right= of A'_out1']{$s$};
 \end{tikzpicture}

%% file: figures/architecture.tex
\begin{tikzpicture}[oriented WD, bbx = .5cm, bby =.8ex, bb min width=.5cm, bb port length=2pt, bb port sep=1]
\node[bb={2}{1}] (P) {{\begin{tabular}{c}
     processor\\ $P_1$
\end{tabular}}};
\node[bb={2}{1},above left= of P] (I) {\begin{tabular}{c}
     IMU\\ $I_1$
\end{tabular}};
\node[bb={2}{1},below left=of P] (J) {\begin{tabular}{c}
     IMU\\ $I_2$
\end{tabular}};
\node[bb={2}{1},inner sep=.1in,fit={($(J)-(3,0)$)(J)(I)(P)},effectivenessCurve,dashed] (L) {};
\node[bb={2}{1}, below right= 1.1cm and 1.5cm of L] (Q) {{\begin{tabular}{c}
     processor\\ $P_2$
\end{tabular}}};
\node[bb={1}{1},right= of Q] (V) {{{\begin{tabular}{c}
     servos\\ $V$
\end{tabular}}}};
\node[bb={2}{1}, inner sep=.1in,fit={(Q)(V)},CostCurve,dashed] (C) {};
\node[bb={1}{1}, above right=0.2cm and 2cm of C] (U) {{\begin{tabular}{c}
     elevator\\ $W$
\end{tabular}}};
\node[bb={1}{1}, above=1em of U] (Z) {{\begin{tabular}{c}
     throttle\\ $Z$
\end{tabular}}};
\node[bb={1}{1}, above=1em of Z] (Y) {{\begin{tabular}{c}
     rudder\\ $Y$
\end{tabular}}};
\node[bb={1}{1}, above=1em of Y] (X) {{\begin{tabular}{c}
     aileron\\ $X$
\end{tabular}}};
\node[bb={4}{1}, right=12.5cm of P] (F) {{\begin{tabular}{c}
     airframe\\ $F$
\end{tabular}}};
\node[bb={1}{1}, fit={($(X.west)-(1,0)$)(X)(U)(F)},yel,dashed] (D) {};
\node[bb={2}{1}, fit={($(L.west)-(1,0)$) ($(D.east)+(1.3,0)$) ($(L.north)+(0,15)$) ($(C.south)-(0,3)$)},gray] (A) {};
\draw[ar] (L_in1') to (I_in1);
\draw[ar] (L_in1') to (J_in1);
\draw[ar] (L_in2') to (J_in2);
\draw[ar] (L_in2') to (I_in2);
\draw[ar] (I_out1) to (P_in1);
\draw[ar] (J_out1) to (P_in2);
\draw[ar] (P_out1) to (L_out1');
 \draw[ar] (A_in2') to (C_in2);
  \draw[ar] (A_in1') to (L_in2);
 \draw[ar] (L_out1) to (C_in1);
 \draw[ar] (C_out1) to (D_in1);
 \draw[ar] (V_out1) to node[above,CostCurve]{$C$} (C_out1');
 \draw[ar] (D_out1) to (A_out1');
\draw[ar] (C_in1') to (Q_in1);
\draw[ar] (C_in2') to (Q_in2);
\draw[ar] (Q_out1) to (V_in1);
\draw[ar] (D_in1') to (X_in1);
\draw[ar] (D_in1') to (Y_in1);
\draw[ar] (D_in1') to (Z_in1);
\draw[ar] (D_in1') to (U_in1);
\draw[ar] (X_out1) to (F_in1);
\draw[ar] (Y_out1) to (F_in2);
\draw[ar] (Z_out1) to (F_in3);
\draw[ar] (U_out1) to (F_in4);
\draw[ar] (F_out1) to (D_out1');
\draw[ar] let \p1=(D.north east), \p2=(L.north west), \n1={\y1+2*\bby}, \n2=\bbportlen in(D_out1) to[in=0] (\x1+\n2,\n1) -- (\x2-\n2,\n1) to[out=180] (L_in1);
\node[below right= .2 and 14 of C]{UAV};
\node[above right= -.5cm and 2.3cm of I,effectivenessCurve]{$L$};
\node[above right= -.5cm and 2.3cm of X,yel]{$D$};
 \end{tikzpicture}